\documentclass{article}
\begin{document}
\newpage
\begin{titlepage}
\begin{flushright}
SLAC-PUB-8434 \\ 
FISIST/5-2000/CFIF\\
April 2000
\end{flushright}
\vspace{0.8cm}
\begin{center}
{\large \bf Mixing and CP violation with Quasidegenerate Majorana 
Neutrinos}
\vspace{1.0cm}

G. C. Branco$^a$ \footnote{e-mail: d2003@beta.ist.utl.pt}
M. N. Rebelo$^{a,b}$ \footnote{
On sabbatical leave, visiting SLAC - Stanford Linear Accelerator 
Center, partially supported by Department of Energy, contract
DE-AC03-76SF00515, FLAD - Funda\c{c}\~{a}o Luso Americana para
o Desenvolvimento and GTAE - Grupo Te\'{o}rico de Altas Energias.\\
E-mail:rebelo@slac.stanford.edu,
rebelo@beta.ist.utl.pt}
J. I. Silva-Marcos$^a$ \footnote{e-mail: juca@gtae2.ist.utl.pt}\\
\medskip
$^a$Centro de F\'{\i}sica das Interac\c{c}\~{o}es Fundamentais (CFIF),\\
Instituto Superior T\'{e}cnico, P-1049-001,
Lisboa, Portugal\\
\medskip
$^b$Stanford Linear Accelerator Center, Stanford University,\\
Stanford, CA  94309
\end{center}
\vglue 0.8truecm


\begin{abstract}
We study the exactly degenerate case for three lefthanded 
Majorana neutrinos building a general parametrization for
the leptonic mixing matrix characterized by two angles and 
one CP violating phase and identify a weak-basis invariant 
relevant, in this case, for CP violation. After  lifting  the
degeneracy, this parametrization accommodates the present data on
atmospheric and solar neutrinos, as well as 
neutrinoless double beta  decay. Some of
the leptonic mixing ans\"atze suggested in the literature 
correspond to
special cases of this parametrization.
\end{abstract}

\vfill
\begin{center}
 To appear in ``Particle Physics on the
Boundary of the Millenium", Proceedings of the 9th Lomonosov 
Conference on Elementary Particle Physics, ed. by Alexander
Studenikin published by Interregional Centre for Advanced 
Studies, Moscow, Russia, 2000
\end{center}

\end{titlepage}
 
\section{Introduction}

In the Standard Model (SM) neutrinos are massless and there 
is no mixing in the leptonic sector. Currently there are indications 
for neutrino oscillations in solar \cite{solar}, atmospheric
\cite{atmos} and accelerator \cite{accel} experiments with the 
strongest evidence coming from the Super-Kamiokande atmospheric 
neutrino data \cite{atmos}. The LSND experiment is the only one
that has not been confirmed independently.  Neutrino oscillations 
are a clear sign of physics beyond the SM requiring the existence
of massive neutrinos. The interpretation of the solar and atmospheric 
anomalies can be done in the framework of three lefthanded neutrinos,
without additional sterile neutrinos. In our work \cite{brs},
we have considered the case of three highly degenerate Majorana 
neutrinos and we analised in detail leptonic mixing and CP violation
in the limit of exactly degenerate masses identifying, in this limit,
a weak-basis invariant which controls the strength of CP violation.
We showed that a two angle parametrization suggested by the exact
degeneracy limit can fit all the present atmospheric and solar
neutrino data in agreement with the experimental bound imposed by 
neutrinoless double beta decay. Our two angle parametrization
has the interesting property of reproducing, for specific choices of 
the angles, several of the mixing schemes proposed in the literature
such as bimaximal mixing \cite{bim}, democratic mixing \cite{dem}
as well as the scheme suggested by Georgi and Glashow \cite{alt}.

\section{The limit of exact degeneracy}
The experimental constraints on squared neutrino mass differences
coming from solar and atmospheric experiments together with the assumption
that neutrino masses might be of the order of $1 eV$
lead to highly degenerate masses \cite{deg}.

The terms of the Lagrangean relevant for our discussion are
\begin{equation}
\label{eq1}{\cal L}_{{{\rm {mass}}}}\ =\ -\ (\nu _{L_\alpha })^T\ C^{-1}\
m_{\alpha \beta \quad }\nu _{L_\beta }\ +{\rm {\ h.c.}} 
\end{equation}
and
\begin{equation}
\label{eq3}{\cal L}_{{{\rm {W}}}}\ =\ \frac g2\ \ \left( 
\begin{array}{ccc}
\overline{e}, & \overline{\mu }, & \overline{\tau } 
\end{array}
\right) _L\gamma _\mu \quad U\quad \left( 
\begin{array}{c}
\nu _1 \\ 
\nu _2 \\ 
\nu _3 
\end{array}
\right) _LW^\mu \quad +{\rm {\ h.c.}} 
\end{equation}
${\cal L}_{{\rm {mass}}}$ is a generic Majorana mass term 
for the three lefthanded neutrinos where
$m=(m_{\alpha \beta })$ is a $3\times 3$ complex symmetric mass
matrix, and $\nu _{L_\alpha }$ denote weak eigenstates. In principle 
the matrix $m$ could be an effective Majorana mass matrix
within a framework with three left-handed and three right-handed neutrinos.
We shall work in the weak-basis (WB) where the charged lepton mass matrix is
diagonal, real and positive. The neutrino mass matrix can be diagonalized by
the transformation
\begin{equation}
\label{eq2}U^T\cdot m\cdot U\ =\ {{\rm {diag}}(m_{\nu _1},\ m_{\nu _2},\
m_{\nu _3})} 
\end{equation}
so that the weak eigenstates $\nu _{L_\alpha }$, are related to the mass
eigenstates, $\nu _{L_i}$, by $\nu _{L_\alpha }=U_{\alpha i}$ $\nu _{L_i}$
and the fields in the charged current given by 
Eq.(\ref{eq3}) are already physical fields.

In general the neutrino diagonalization matrix $U$ considered above
can be parametrized by three angles and three phases that are CP
violating. It is well known that in the case of Dirac neutrinos
only three angles and one phase are required to parametrize 
$U$ and that in the limit of exact degeneracy $U$ can be rotated away 
into the unit matrix through a redefinition of the neutrino fields.
In this respect Majorana neutrinos
are fundamentally different and  only if the theory is CP 
invariant and the three degenerate neutrinos have the same 
CP parity, can the matrix 
$U$ be rotated away.

In the limit of exact degeneracy with $\mu $ the common neutrino mass
we shall denote the mixing matrix by $U_{\circ }$ and define a
 dimensionless mass matrix by  $Z_{\circ }$ given by $Z_{\circ }=\
m/\mu $. From Eq(\ref{eq2}) we obtain 
\begin{equation}
\label{eq4}Z_{\circ }\ =U_{\circ }^{\star }\cdot U_{\circ }^{\dagger }\ 
\end{equation} 
so that $Z_{\circ }$ is a unitary symmetric matrix and can be written 
without loss of generality as
\begin{equation}
\label{eqex}Z_{\circ }\ =\ K^{*} \cdot  
\left( 
\begin{array}{ccc}
c_\theta & s_\theta c_\phi & s_\theta s_\phi \\
s_\theta c_\phi & w_{1} & w_{2} \\ 
s_\theta s_\phi & w_{2} & w_{3} 
\end{array}
\right) \cdot K^{*} 
\end{equation}
where $c$ and $s$ stand for cosine and sine respectively,
the $w_i$ may be complex entries and $K$ is a diagonal unitary matrix .
After a WB transformation under which $Z_{\circ }\rightarrow \
K\cdot Z_{\circ }\cdot K$, $Z_{\circ }$ transforms into
\begin{equation}
\label{eq5}Z_{\circ }\ =\ \left( 
\begin{array}{ccc}
1 & 0 & 0 \\ 
0 & c_\phi & s_\phi \\ 
0 & s_\phi & -c_\phi 
\end{array}
\right) \cdot \left( 
\begin{array}{ccc}
c_\theta & s_\theta & 0 \\ 
s_\theta & z_{22} & z_{23} \\ 
0 & z_{23} & z_{33} 
\end{array}
\right) \cdot \left( 
\begin{array}{ccc}
1 & 0 & 0 \\ 
0 & c_\phi & s_\phi \\ 
0 & s_\phi & -c_\phi 
\end{array}
\right) 
\end{equation}
Unitarity of $Z_{\circ }$, requires that either $s_\theta $ or $z_{23}$
vanish. The case $s_\theta =0$
leads to CP invariance. Assuming $s_\theta \neq 0$ the
most general form for $Z_{\circ }$ is given by: 
\begin{equation}
\label{eq6}Z_{\circ }\ =\ \left( 
\begin{array}{ccc}
1 & 0 & 0 \\ 
0 & c_\phi & s_\phi \\ 
0 & s_\phi & -c_\phi 
\end{array}
\right) \cdot \left( 
\begin{array}{ccc}
c_\theta & s_\theta & 0 \\ 
s_\theta & -c_\theta & 0 \\ 
0 & 0 & e^{i\alpha } 
\end{array}
\right) \cdot \left( 
\begin{array}{ccc}
1 & 0 & 0 \\ 
0 & c_\phi & s_\phi \\ 
0 & s_\phi & -c_\phi 
\end{array}
\right) 
\end{equation}
In the cases $\alpha = 0, \pi$ there is, once again, CP conservation.
It can be readily verified that this parametrization does not include
the trivial case where CP is a good symmetry and all neutrinos have the
same CP parity \cite{wol},
corresponding to the eigenvalues $(1,1,1)$. 
In fact in the case of CP consevation $Z_{\circ }$ is 
diagonalized by an orthogonal transformation leaving invariant both
${\rm {Tr}}(Z_{\circ })$ and  $\det (Z_{\circ})$ and there is no 
choice of parameters in  Eq.(\ref{eq6}) leading to a trace and 
determinant corresponding to this particular case.
On the other
hand the set of eigenvalues $(1, -1, 1)$ and $(1, -1, -1)$ corresponding
to one neutrino with opposite CP parity to the other two can be obtained
for $\alpha=0 $ and $\alpha =\pi $, respectively.
The diagonalization of the matrix  $Z_{\circ }$ through the 
transformation of Eq.(\ref{eq2}), together with the requirement 
of positive diagonal elements requires a $U_{\circ }$ matrix given by
\begin{equation}
\label{eq7}U_{\circ }\ =\ \left( 
\begin{array}{ccc}
1 & 0 & 0 \\ 
0 & c_\phi & s_\phi \\ 
0 & s_\phi & -c_\phi 
\end{array}
\right) \cdot \left( 
\begin{array}{ccc}
\cos (\frac \theta 2) & \sin (\frac \theta 2) & 0 \\ 
\sin (\frac \theta 2) & -\cos (\frac \theta 2) & 0 \\ 
0 & 0 & e^{-i\alpha /2} 
\end{array}
\right) \cdot \left( 
\begin{array}{ccc}
1 & 0 & 0 \\ 
0 & i & 0 \\ 
0 & 0 & 1 
\end{array}
\right) 
\end{equation}
$U_{\circ }$ is not an orthogonal matrix even in the CP conserving
limit, i.e., for $\alpha =0$, $\pi $, as a result it cannot be rotated 
away through a redefinition of the neutrino fields.

From this discussion we see that in 
the case of three degenerate Majorana neutrinos the parametrization
of $U$ requires two angles and one phase and there may be 
CP violation \cite{xyz}.It can be
shown that a necessary and sufficient condition for CP
invariance, in the degenerate limit, is: 
\begin{equation}
\label{eq8}G\equiv \ {\rm {Tr}}\left[ \ (m\cdot h\cdot m^{\star })\ ,\
h^{\star }\right] ^3\ =\ 0 
\end{equation}
where $h=m_\ell \cdot m_\ell ^{\dagger }$, and $m_\ell $ denotes the charged
lepton mass matrix. $G$ is a WB invariant and can be written as
\begin{equation}
\begin{array}{l}
\label{eq9}G=6i\ \Delta _m~{\rm {Im}}[(Z_{\circ })_{11}(Z_{\circ
})_{22}(Z_{\circ })_{12}^{\star }(Z_{\circ })_{21}^{\star }] \\
=\frac{3i}2\
\Delta _m~\cos (\theta )\sin ^2(\theta )\ \sin ^2(2\phi )\ \sin (\alpha )
\end{array} 
\end{equation}
where $\Delta _m=$\ $\mu ^6\ (m_\tau ^2-m_\mu ^2\ )^2(m_\tau ^2-m_e^2\
)^2(m_\mu ^2-m_e^2\ )^2$ is a multiplicative factor which contains the
different masses of the charged leptons and the common neutrino mass $\mu $.
In Refs.\cite{xyz} \cite{rst} various examples of 
CP-odd WB-invariants were
constructed, but all of those invariants automatically vanish in the limit
of exact degeneracy whilst the invariant of  Eq.(\ref{eq8}) only
vanishes if CP is conserved. It was shown in Ref. \cite{brs} that 
the imposition of maximal CP violation leads to a
structure of the Majorana neutrino mass of the type that one obtains in the
framework of universal strength for Yukawa couplings $\cite{ref7a}$.

\section{Lifting the degeneracy. Phenomenological implications}
In the limit of exact
degeneracy, the leptonic mixing matrix $U_{\circ}$
is param- etrized by two angles 
$\theta $, $\phi $ and one phase $\alpha $ and is defined 
only up to an arbitrary orthogonal transformation 
$U_{\circ }\rightarrow U_{\circ}\cdot O$. The physically interesting case
corresponds to quasidegenerate neutrinos. Let us assume that the
degeneracy is lifted through a small perturbation:
\begin{equation}
\label{eq10}Z=Z_{\circ }+\varepsilon \ Q\ \ 
\end{equation}
where $\varepsilon $ is a small parameter and $Q$ is a symmetric complex
matrix of order one. It was shown in Ref. \cite{brs} that 
in the presence of a small perturbation around the degeneracy limit,
the mixing matrix becomes, to leading order, $U_{\circ }\cdot O$,
where $O\,$ is no longer arbitrary, being the orthogonal matrix which
diagonalizes the symmetric real matrix $A$, defined by 
$A={\rm {Re}}(U_{\circ }^T \cdot Q\cdot U_{\circ })$. As a result,
for quasidegenerate neutrinos, to leading order, only one CP 
violating phase appears in the leptonic mixing,
namely the phase $\alpha $ present in $U_{\circ }$ and the question
of whether the two angle parametrization given by 
Eq.(\ref{eq7}) can accommodate the present experimental data on
atmospheric and solar neutrinos, as well as the constraints 
on double beta decay immediately arises. Of course this corresponds 
to the case where the matrix
$U_{\circ }^T \cdot Q\cdot U_{\circ }$
is already a real and diagonal matrix lifting the 
degeneracy as required experimentally.

The constraints arising from neutrinoless double beta decay 
put an upper bound on $<m>$, an average neutrino mass, given
in standard notation by
\begin{equation}
\label{eq13}<m>\ =\ {\sum}_i\ U_{ei}^2\ m_{\nu _i}\ =\ m_{ee}^{\star }\ \ 
\end{equation}
where the $U_{ei}$ denote the elements of the first row of the mixing matrix 
$U$, and $m_{ee}$ is the $(1,1)$ element of the mass matrix $m$.
At present, the strongest bound is 
$|<m>|=|m_{ee}|<0.2\ eV$ $\cite{ref8}$.
In the limit of exact degeneracy, we
have $m_{ee}=\mu \ \cos (\theta )$, where we have used the parametrization
of Eq.(\ref{eq7}). If we fix $\mu =1\ eV$, then neutrino masses are equal to
a precision sufficient to neglect their differences, and the experimental
bound on $m_{ee}$ immediately translates into a single bound on the
parameter $\theta $, namely $|\cos (\theta )|<0.2$.

In our framework, without sterile neutrinos, the
atmospheric neutrino data
supports the existence of oscillations of atmospheric neutrinos to tau
neutrinos with a large mixing angle satisfying the
bound $\sin ^2(2\theta _{{\rm {atm}}})>0.82$, and the neutrino mass square
difference in the range $5\times 10^{-4}\ eV^2<\Delta m_{{\rm {atm}}
}^2<6\times 10^{-3}\ eV^2$. This interpretation is further supported by
recent data from  the CHOOZ 
reactor netrino experiment \cite{cho} leading to the upper bound 
$\mid U_{e3}\mid \leq (0.22-0.14)$.

In the context of three left-handed neutrinos, the probability for a
neutrino $\nu _\alpha $ to oscillate to other neutrinos is
\begin{equation}
\label{eq13a}1-P(\nu _\alpha \rightarrow \nu _\alpha )=4\ {\sum }_{i<j}\
U_{\alpha i}U_{\alpha i}^{\star }U_{\alpha j}^{\star }U_{\alpha j}\quad \sin
^2\left[ \frac{\Delta m_{ji}^2}4\frac LE\right] 
\end{equation}
where $\Delta m_{ji}^2=|m_j^2-m_i^2|$, $E$ is the neutrino energy and $L$
denotes the distance travelled by the neutrino between the source and the
detector. Since in the range $L/E$ that is relevant for atmospheric
neutrinos the term in $\sin ^2[(\Delta m_{21}^2/4)(L/E)]$ can be
disregarded, we may identify $\sin ^2(2\theta _{{\rm {atm}}})$ with 
$4(U_{21}U_{21}^{\star }U_{23}^{\star }U_{23}+$ $ U_{22}U_{22}^{\star
}U_{23}^{\star }U_{23})$.\ In the framework of our two-angle parametrization
of Eq.(\ref{eq7}), the above combination of matrix elements has a simple
form and one obtains $\sin ^2(2\theta _{{\rm {atm}}})=\sin ^2(2\phi )$,
i.e., $\theta _{{\rm {atm}}}$ can be identified with the angle $\phi $ and
thus the atmospheric neutrino data leads to the constraint $\sin ^2(2\phi
)>0.82$.

The discrepancy between the observed and the calculated $\cite{ref11}$ solar
neutrino fluxes also requires neutrino oscillations, although at this stage
various schemes are still possible, namely within the framework of the MSW
mechanism $\cite{ref12}$ there is a small angle solution $\sin ^2(2\theta _{%
{\rm {sol}}})\approx 7\times 10^{-3}\ $ with $\Delta m_{{\rm {sol}}%
}^2\approx 6\times 10^{-6}\ eV^2$, and a large angle solution $\sin
^2(2\theta _{{\rm {sol}}})\sim 0.65-0.97$ with $\Delta m_{{\rm {sol}}%
}^2\sim (2-20) \times 10^{-5}\ eV^2$. 
Another solution could be vacuum
oscillations with $\sin ^2(2\theta _{{\rm {sol}}})\approx 0.9$ and $\Delta
m_{{\rm {sol}}}^2\approx 10^{-10}\ eV^2$. Since in our two-angle
parametrization one has $U_{13}=0$ we obtain $\sin ^2(2\theta _{{\rm {sol}}%
})=$ $4U_{11}U_{11}^{\star }U_{12}^{\star }U_{12}$ leading to $\sin
^2(2\theta _{{\rm {sol}}})=\sin ^2(\theta )$, i.e., in our parametrization $%
2\theta _{{\rm {sol}}}=\theta $.

We see that each of the experiments considered above independently 
constrains a single parameter. Also, it is clear from 
Eq.(\ref{eq13}), that with small solar neutrino mixing the
bound from double beta decay would not be satisfied for
quasidegenerate neutrinos with masses of the order of $1\  eV$.

Finally we show that
some of the neutrino mixing schemes proposed in the
literature correspond to specific cases of the two-angle parametrization
suggested by Eq.(\ref{eq7}).

(a) {\it Bimaximal Mixing} $\cite{bim}$: In this scheme the lines of the
neutrino mixing matrix have the following structure: 
\begin{equation}
\label{eq14}
\begin{array}{ccc}
L_1=\left( \frac 1{\sqrt{2}},\frac{-1}{\sqrt{2}},0\right) ;\quad & 
L_2=\left( \frac 12,\frac 12,\frac 1{\sqrt{2}}\right) ;\quad & L_3=\left( 
\frac{-1}2,\frac{-1}2,\frac 1{\sqrt{2}}\right) 
\end{array}
\end{equation}
This pattern of neutrino mixing is obtained within the two-angle
parame-
trization for the following values of $\theta $, $\phi $ and $\alpha $%
: 
\begin{equation}
\label{eq15}\alpha =0\ ;\quad \cos (\theta /2)=-\sin (\theta /2)=-\cos (\phi
)=\sin (\phi )=\frac 1{\sqrt{2}} 
\end{equation}

(b) {\it Democratic Mixing} $\cite{dem}$: 
In this case the neutrino mixing matrix has, to a
very good approximation, the following form: 
\begin{equation}
\label{eq16}
\begin{array}{ccc}
L_1=\left( \frac 1{\sqrt{2}},\frac{-1}{\sqrt{2}},0\right) ;\quad & 
L_2=\left( \frac 1{\sqrt{6}},\frac 1{\sqrt{6}},\frac{-2}{\sqrt{6}}\right)
;\quad & L_3=\left( \frac 1{\sqrt{3}},\frac 1{\sqrt{3}},\frac 1{\sqrt{3}%
}\right) 
\end{array}
\end{equation}
Within the two-angle parametrization, one obtains the democratic mixing for
the following values of the parameters: 
\begin{equation}
\label{eq17}\alpha =0\ ;\quad \cos (\theta /2)=-\sin (\theta /2)=\frac 1{%
\sqrt{2}}\ ;\quad \cos (\phi )=\frac 1{\sqrt{2}}\sin (\phi )=\frac{-1}{\sqrt{%
3}} 
\end{equation}

In the above analysis, we have not paid attention to the factors ``$i$''
appearing in our two-angle parametrization of Eq.(\ref{eq7}). As we have
previously emphasized, these factors of ``$i$'' have to do with the fact
that in the construction of the two-angle parametrization, we have
implicitly assumed that in the limit of CP\ invariance (i.e. $\sin (\alpha
)\rightarrow 0$), one of the Majorana neutrinos has relative CP parity
opposite to the other two. The factors of ``$i$'' do not play any r\^ole in
the analysis of atmospheric and solar neutrino data, but are crucial in the
analysis of double beta decay.

(c) {\it Georgi-Glashow mass matrix} $\cite{alt}$: Using an analysis of the
present neutrino data Georgi and Glashow have suggested the following
approximate form for the Majorana neutrino mass matrix 
\begin{equation}
\label{eq18}
\begin{array}{l}
(m)_{1i}=\mu \left( 0,\frac 1{\sqrt{2}},\frac 1{\sqrt{2}}\right) ;\quad 
(m)_{2i}=\mu \left( \frac 1{\sqrt{2}},\frac 12,\frac{-1}2\right) ; \\ 
(m)_{3i}=\mu \left( \frac 1{\sqrt{2}},\frac{-1}2,\frac 12\right) 
\end{array}
\end{equation}
From Eq.(\ref{eq6}) it follows that this neutrino mass matrix is obtained,
within the two-angle parametrization for the following values of its
parameters, 
\begin{equation}
\label{eq19}
\begin{array}{ccc}
\alpha =0; & \sin (\theta )=1; & \cos (\phi )=\sin (\phi )=\frac 1{\sqrt{2}} 
\end{array}
\end{equation}

\section*{Acknowledgments}
M. N. R is grateful to the Organizers of the 
the 9th Lomonosov Conference on Elementary Particle
Physics for the warm hospitality. M. N. R. received financial
support from FCT - Funda\c{c}\~{a}o de Ci\^{e}ncia e Tecnologia
through GTAE - Grupo Te\'{o}rico de Altas Energias and 
CFIF - Centro de F\'{\i}sica das Interac\c{c}\~{o}es Fundamentais,
to participate at the Conference.


\end{document}